\documentclass{aa}  

\usepackage{graphicx}
\usepackage{color}
\usepackage{subfigure}
\usepackage{longtable}
\usepackage{multicol}
\usepackage{placeins}

\usepackage{txfonts}
\begin{document}

   \title{Decoding NGC 7252 as a blue elliptical galaxy}

  \author{Koshy George
          \inst{1} }
  \institute{$^{1}$Faculty of Physics, Ludwig-Maximilians-Universit{\"a}t, Scheinerstr. 1, 81679, Munich, Germany\\
                \email{koshyastro@gmail.com}}


 

  \abstract
  {Elliptical galaxies with blue optical colours and significant star formation are hypothesised to be major merger remnants of gas-rich spiral galaxies or normal elliptical galaxies with a sudden burst of star formation. We present here a scenario in which blue elliptical galaxies identified in shallow imaging surveys may fail to recover faint features that are indicative of past merger activity using a nearby major merger remnant. Based on deep optical imaging data of the post-merger galaxy, NGC 7252, we demonstrate that the galaxy can appear as an elliptical galaxy if it is observed at higher redshifts. The main body and the low surface brightness merger features found at the outskirts of the galaxy are blue in the optical $g-r$ colour map. We argue that the higher-redshift blue elliptical galaxies discovered in surveys as shallow as the SDSS or DECaLS may be advanced mergers whose defining tidal features fall below the detection
limits of the surveys. This should be taken into consideration during the morphological classification of these systems in future and ongoing surveys.}


   \maketitle
%

\section{Introduction}

Massive galaxies in the local Universe are morphologically classified as elliptical (E), S0, and spirals. E/S0 galaxies are generally observed to be gas poor, without significant ongoing star formation. Spiral galaxies, on the other hand, tend to be gas rich and are actively forming stars. This is reflected as distinct bimodal regions in the optical colour-magnitude diagram and in diagrams of the stellar mass-star formation rate \citep{Baldry_2004,Brinchmann_2004,Salim_2007,Noeske_2007,Elbaz_2007,Daddi_2007}. The bimodal nature of star formation in galaxies could be explained by internal or external processes due to which star-forming galaxies cease to form new stars. This could be associated with a morphological change through which a star-forming spiral galaxy can become a non-star-forming E/S0 galaxy. This is possible through major mergers, in which two equal-mass star-forming spiral galaxies merge to form a massive elliptical galaxy. In an environmental process, star-forming spiral galaxies can also fall into galaxy clusters and groups without subsequent supply of gas, which halts star formation and transforms the morphology into an S0 galaxy. This is supported by the redshift evolution of stellar mass buildup in red-sequence (E/S0) galaxies that occurs at the expense of blue cloud (spiral) galaxies \citep{Bell_2004,Faber_2007,Brown_2007}. However,  the star formation rates of a small fraction of blue colour E/S0 galaxies are as
significant as in spiral galaxies \citep{Fukugita_2004,Schawinski_2009,Kannappan_2009,Huertas_2010,McIntosh_2014,Mahajan_2018,Moffett_2019,Dhiwar_2022,Paspaliaris_2023,Lazar_2023}. These galaxies are found to be in low-density regions with blue optical colours, and they occupy the main sequence of star-forming galaxies. The formation of blue elliptical galaxies is hypothesised as follows. Two gas-rich, equal-mass spiral galaxies can merge to form an elliptical galaxy with significant star formation (this is one of the channels for elliptical galaxy formation). The star formation will cease when the gas is exhausted, and the galaxy will change to a normal elliptical galaxy. The other scenario involves a rejuvenation process in which the normal elliptical galaxy acquires much gas that then collapses to form new stars.\\

\begin{figure*}
\centering
\includegraphics[width=7in]{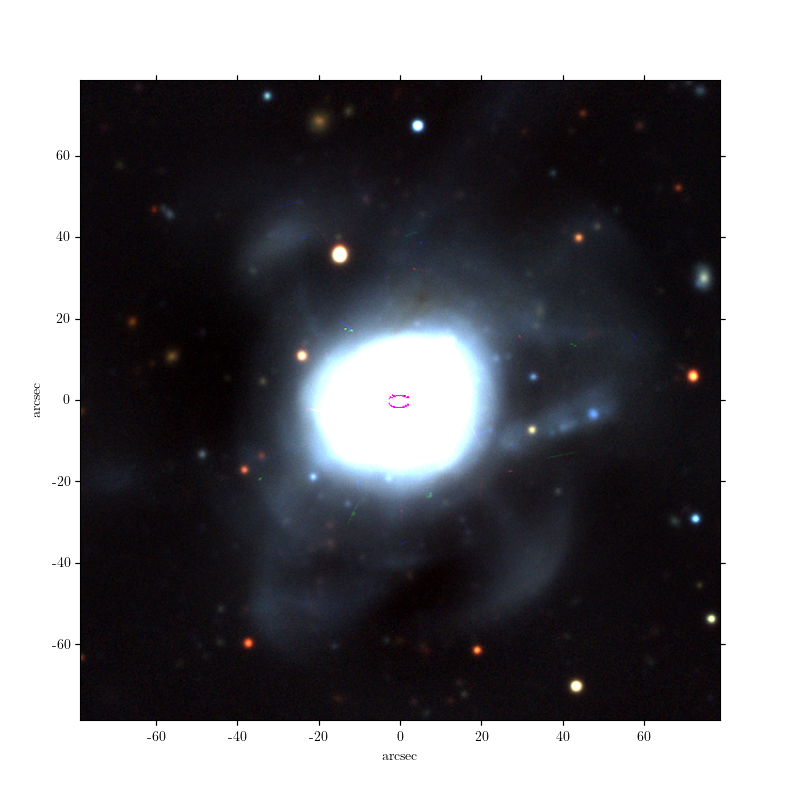}
\caption{Optical colour-composite image of NGC 7252 made from legacy survey $g,r,z$ filter bandpass images (blue, green, and red). The image size is 2.6 $\arcmin$ $\times$ 2.6 $\arcmin$ ($\sim$ 50 kpc $\times$ 50 kpc).} \label{figure:fig1}
\end{figure*}

We test the hypothesis of a major merger origin for blue elliptical galaxies using the known major merger remnant NGC 7252. The galaxy main body is a single-nucleus merger remnant and represents the final stages of the Toomre sequence of merging, where the merger remnant will eventually deplete the fuel for star formation and evolve into an elliptical galaxy \citep{Toomre_1972,Toomre_1977,Schweizer_1982}. The advanced merger between two gas-rich spiral galaxies results in tidal tails, shells, and ripples around the main body in optical imaging \citep{Schweizer_1982,Dupraz_1990,Wang_1992,Fritze_1994,Hibbard_1994}. Significant star formation is detected in the main body and at the outskirts of the galaxy, with indications of a gaseous disk and possible active galactic nucleus (AGN) feedback at the centre \citep{George_2018a,Weaver_2018,George_2018b}. The merger and the associated starburst in the galaxy are understood to have started 600–700 Myr ago \citep{Hibbard_1995,Chien_2010}. The surface brightness profile of the galaxy main body follows a de Vaucouleurs profile (typical of elliptical galaxies), in which optical spectroscopy reveals post-starburst features \citep{Schweizer_1982,Hibbard_1999}. The galaxy follows the scaling relations of normal elliptical galaxies, such as the Faber–Jackson and fundamental plane relation \citep{Lake_1986, Hibbard_1995,Genzel_2001,Rothberg_2006}. We used deep optical imaging data of NGC 7252 to investigate whether the galaxy shows properties similar to that of blue elliptical galaxies. Blue elliptical galaxies could have formed from a similar equal-mass spiral galaxy merger, in which the merger features reach beyond the detection limit of the wide-field optical surveys  (The Sloan Digital Sky Survey (SDSS) and The Dark Energy Camera Legacy Survey (DECaLS)) based on which these galaxies are originally classified. We explore the optical $g-r$ colour of the low surface brightness features in the outskirts and compare the values against the main body of the galaxy. We place the galaxy at various redshifts as imaged at the surface brightness limit of the legacy survey imaging used in this work by correcting for angular size distance and applying cosmological surface brightness dimming. We adopt a flat Universe cosmology with $H_{\rm{o}} = 71\,\mathrm{km\,s^{-1}\,Mpc^{-1}}$, $\Omega_{\rm{M}} = 0.27$, $\Omega_{\Lambda} = 0.73$ \citep{Komatsu_2011}.\\

\section{Data and analysis}

\begin{figure}
\centering
{\includegraphics[width=0.52\textwidth]{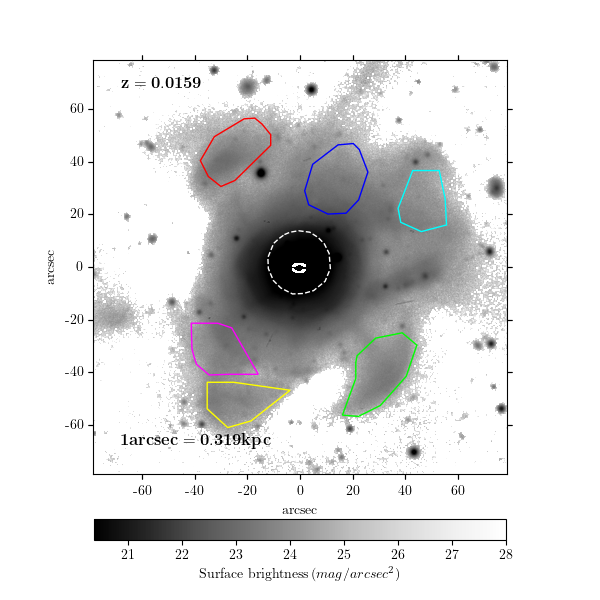}}
\caption{Surface brightness map of NGC 7252 made from $r$-band imaging data. The integrated surface brightness is computed from selected regions along the low surface brightness merger features marked in differently coloured polygons. The galaxy main body is marked with a white outline.} \label{figure:fig2a}
\end{figure}

\begin{figure*}
\centering
{\includegraphics[width=0.43\textwidth]{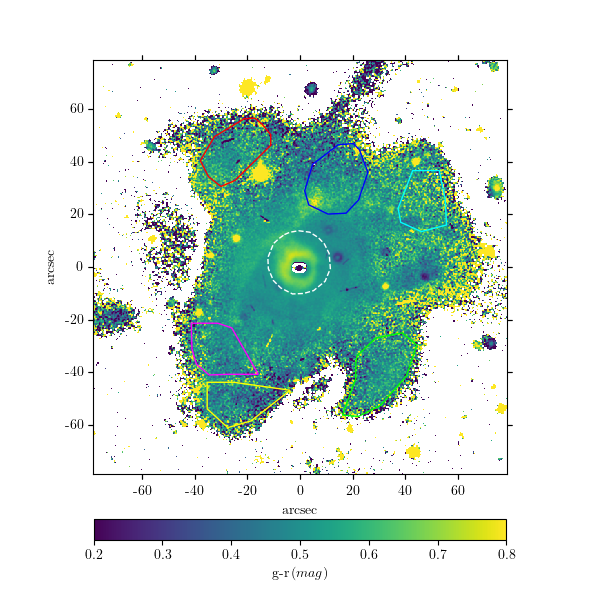}}
{\includegraphics[width=0.56\textwidth]{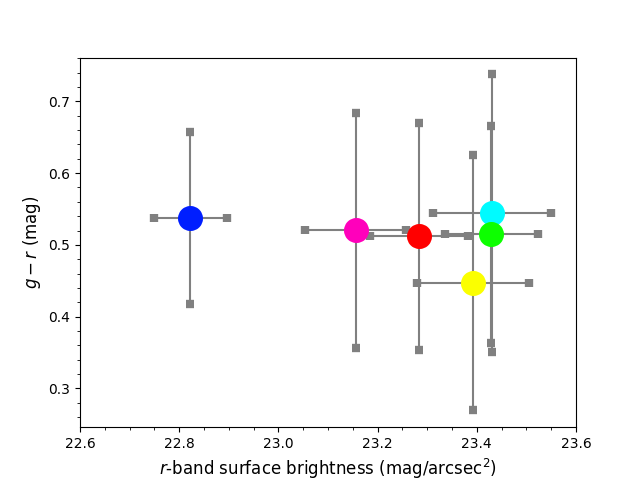}}
\caption{$g-r$ color map of NGC 7252. The $g-r$ colour is computed from selected regions along the low surface brightness merger features marked in differently coloured polygons. The galaxy main body is marked with a white outline. The $r$-band surface brightness from selected regions is plotted against the $g-r$ colour. The colour scheme of points is the same as in the selected regions.} \label{figure:fig2b}
\end{figure*}

NGC 7252 (RA:22:20:44.7,Dec:24:40:42) is a nearby major merger remnant with a spectroscopic redshift (z) $=$ 0.0159 \citep{Rothberg_2006}. The optical $g,r,z$-band imaging data of NGC 7252 were taken from Data Release 10
of the legacy survey DECaLS \citep{Dey_2019}. DECaLS uses the Dark Energy Camera, consisting of 62 2k $\times$ 4k CCDs with a pixel scale of 0.262 arcsec pix$^{-1}$ and a 3.2 deg${^2}$ field of view at the 4 m Blanco telescope of Cerro Tololo Inter-American Observatory. The observations are conducted in a dynamic observing mode in which the exposure times and target field selection can be changed on the fly, depending on the observing conditions to ensure a uniform depth across the survey. The median full width at half maximum (FWHM) in the $g,r,z$ band of the delivered image quality is $\sim$ 1.3, 1.2, and 1.1 arcsec. The  photometric calibration was made using the Pan-STARRS1 DR1 photometry through the set of colour transformation equations given in \citet{Dey_2019}. The $g,r,z$ band coadded images we used in our analysis were calibrated with pixel values stored in nannomaggies, which can be converted into magnitudes using the appropriate conversion stored in the header. DECaLS imaging reaches $\sim$ 2 mag deeper than that of the SDSS and hence can detect low surface brightness features in the r band down to 28 mag arcsec$^{-2}$ (the corresponding limit for SDSS is 25 mag arcsec$^{-2}$) \citep{Driver_2016, Hood_2018}. \\

We used the $grz$ imaging data to create a colour-composite image of NGC 7252 by assigning blue ($g$), green ($r$), and red ($z$) colours. Figure~\ref{figure:fig1} shows the $grz$ colour-composite image of NGC 7252. We note that the pixels from the central region of the galaxy main body are saturated in $r$-band imaging data. The region covered by saturated pixels was masked and was not used for further analysis. Faint tidal features from the recent merger activity are detected around the galaxy. The morphological features around the galaxy were evaluated using the $r$-band imaging. The surface brightness map was created from the $r$-band imaging and is shown in Figure~\ref{figure:fig2a} with an inverted greyscale to detect faint low surface brightness features. We attempted to bring out the faint features by smoothing the pixel noise through running a Gaussian of $\sigma$=1. We visually selected different regions by avoiding foreground stars (and likely stellar clusters) outside the galaxy that we indicated with colour polygons, and the galaxy main body is marked with a white contour. The optical $g-r$ color map of the galaxy was created from the flux-calibrated $g,r$ coadded images ($g-r$ = -2.5 $\times$ log$_{10}$(flux$_{g}$/flux$_{r}$)) and is shown in Figure~\ref{figure:fig2b}. The $g,r$ surface brightness was computed for the marked regions of faint features in the outskirts and the galaxy main body. The $r$-band surface brightness is plotted against the $g-r$ colour of the regions and is shown in Figure~\ref{figure:fig2b}. The selected merger features around the galaxy have integrated blue colours with a median $g-r$ = 0.52. The main body of the galaxy has an early-type morphology with $g-r$ colour = 0.42. The colour value should be treated as a lower limit estimate as the central region saturated pixels in $r$-band imaging are masked and were not used to compute the colour.\\

\begin{figure*}
\centering
{\includegraphics[width=0.24\textwidth]{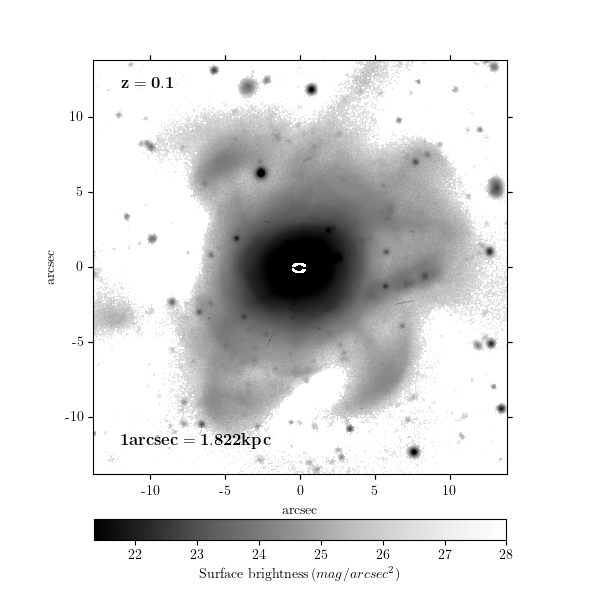}}
{\includegraphics[width=0.24\textwidth]{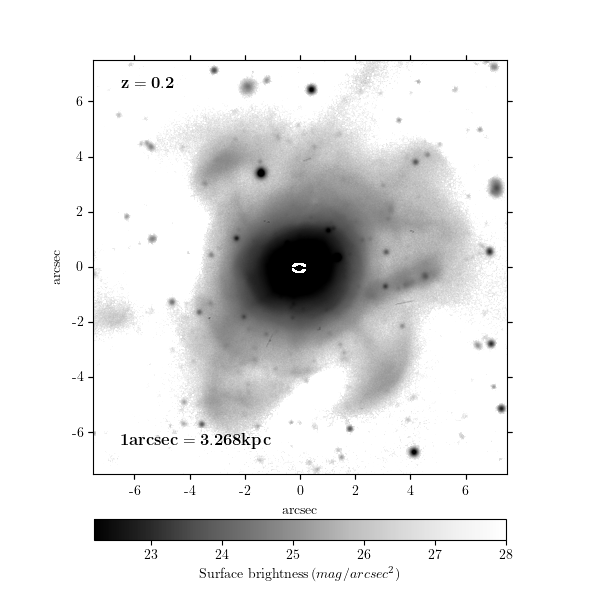}}
{\includegraphics[width=0.24\textwidth]{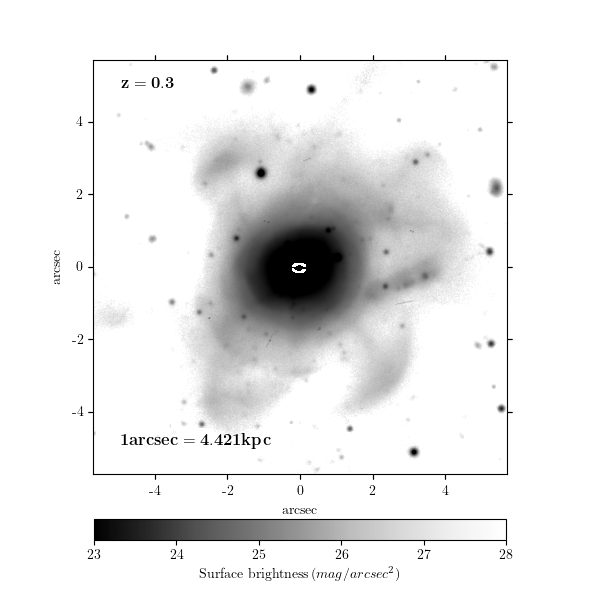}}
{\includegraphics[width=0.24\textwidth]{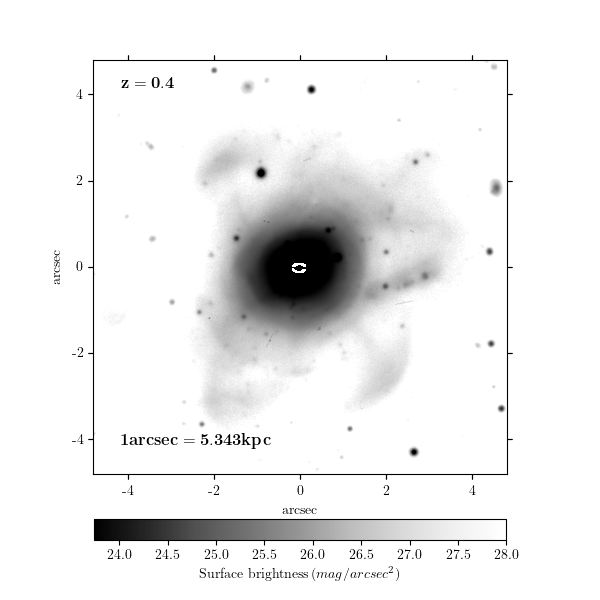}}
{\includegraphics[width=0.24\textwidth]{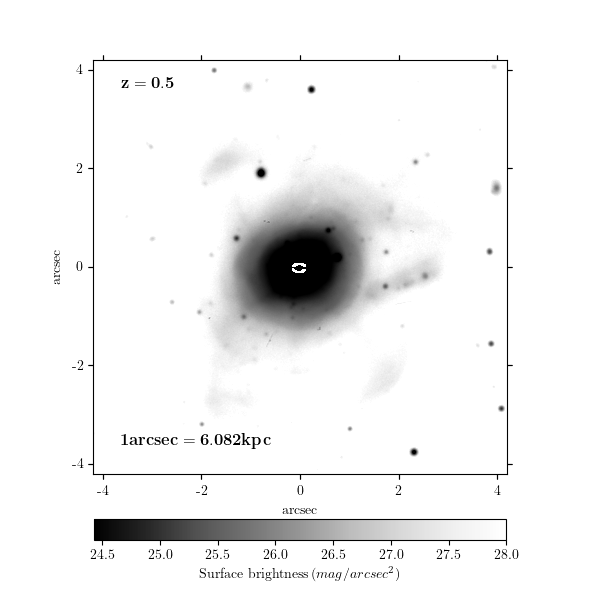}}
{\includegraphics[width=0.24\textwidth]{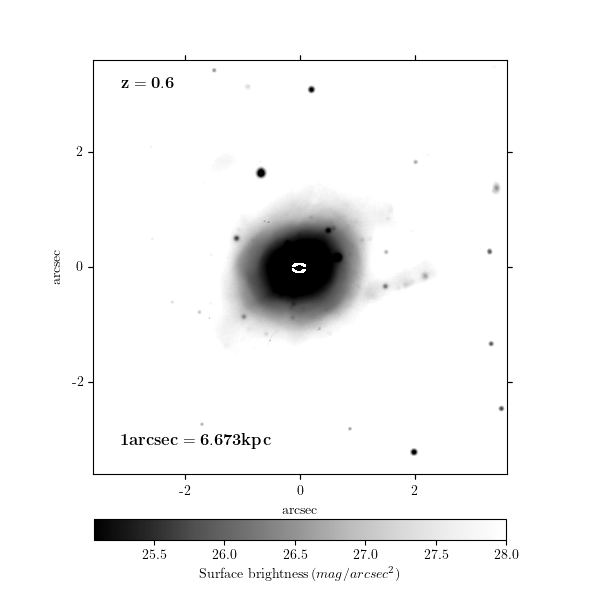}}
{\includegraphics[width=0.24\textwidth]{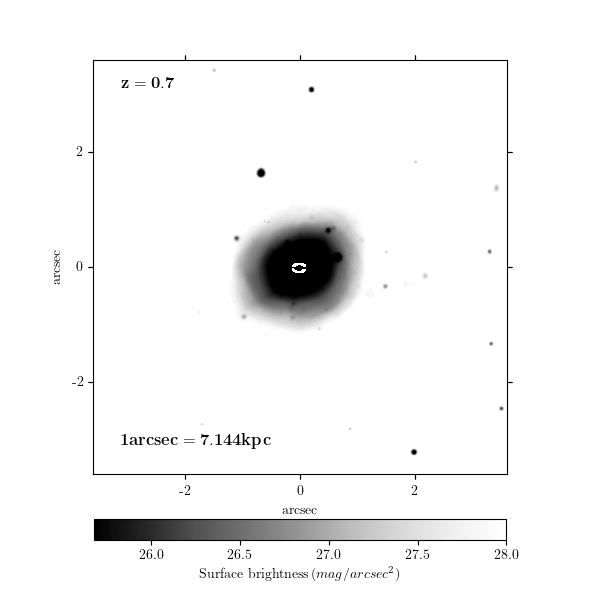}}
{\includegraphics[width=0.24\textwidth]{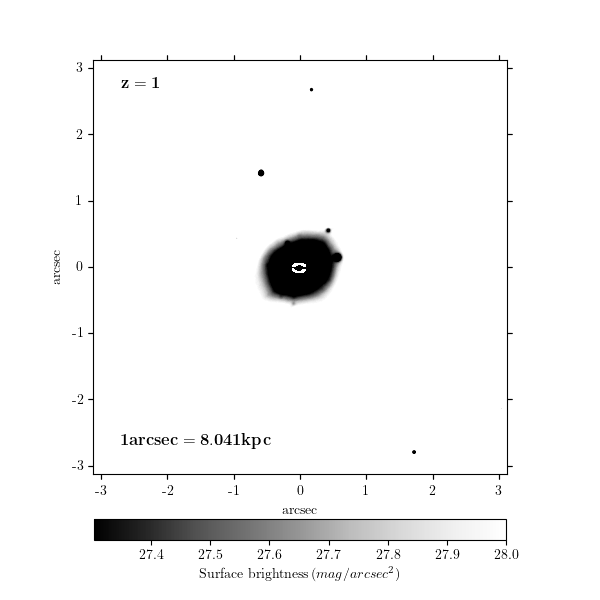}}
\caption{Surface brightness map of NGC 7252 made from $r$-band imaging data as it could have appeared in different redshifts. The corresponding 1$\arcsec$ to kiloparsec conversion for each redshift is given inside the plots. A movie version of the plots is available online.}\label{figure:fig3}
\end{figure*}

We now investigate the appearance of the galaxy at different redshifts in increments of 0.1 up to z=1. We changed the galaxy size for the changing angular size distance with redshift and also took the effect of the cosmological surface brightness dimming ($\mu + 10 \times log_{10}(1+z)$) on the surface brightness map ($\mu$) of the galaxy into account. The $r$-band surface brightness maps for different redshifts between $0>z>1$ are shown in Figure~\ref{figure:fig3}. We limited the surface brightness to 28 mag/arcsec${^2}$ at the detection limit of the legacy survey for every redshift. The surface brightness values that reach beyond this limit are not shown and will not be detected at that redshift. This is the expected $r-band$ appearance of NGC 7252 at different redshifts when observed with a 4m telescope. We note that this is a very simplified scenario that we put forward here for galaxies with a flat spectral energy distribution. The observed $r-band$ in reality will be receiving photons emitted from the rest frame $u-band$ at z $\sim$ 1. \\

\begin{figure}
\centering
\includegraphics[width=3.8in]{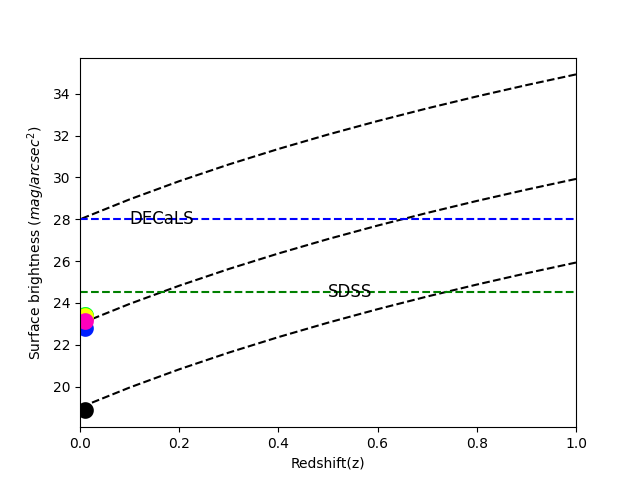}
\caption{Surface brightness variation with redshift due to the cosmological dimming (dotted black line). The surface brightness detection limits from the SDSS and DECaLS sky surveys are shown as green and blue lines. The integrated surface brightness of the main body of the galaxy is shown with a black point. The $r$-band surface brightness from selected regions at the outskirts is shown as coloured points, as in Figure~\ref{figure:fig2a}.} \label{figure:fig4}
\end{figure}

We show the surface brightness change as a function of redshift due to the cosmological surface brightness dimming in Figure~\ref{figure:fig4}. The dotted black lines show the change in three different values of $\mu$ from z =0 to 1. Limiting surface brightness for SDSS and DECaLS surveys are shown by blue and green horizontal lines. The black point is the measured surface brightness for the main body of NGC 7252 and the coloured points are the surface brightness of the faint merger features seen at outskirts as defined in Figure~\ref{figure:fig2a}.\\

\section{Discussion}

The $\Lambda$ cold dark matter paradigm predicts that elliptical  galaxies are formed through a hierarchical merging scenario \citep{DeLucia_2006}. Multiple mergers involving different mass scales are possible, and the gas content  dictates the star formation properties of the merger remnant. NGC 7252 is a nearby major merger remnant understood to have formed from a recent ($<$ 700 Myr) equal-mass merger between two gas-rich spiral galaxies that created the main body around which tidal tails, shells, and ripples are formed, as shown in the optical colour-composite image in Figure~\ref{figure:fig1}. The main body of the galaxy shows properties typical of
elliptical galaxies. The galaxy, however, resides in the blue cloud of the galaxy colour-magnitude diagram \citep{Weaver_2018}, which is at odds with the normal elliptical galaxies. The stellar mass for NGC 7252 is computed to be $\sim$ 10$^{10.6}$ M$\odot$ \citep{Weaver_2018}. \\

Optical imaging data from wide-field surveys such as  SDSS with an integration time $\sim$ 54sec were used to classify blue E/S0 galaxies based on morphology \citep{Fukugita_2004,Schawinski_2009}. \citet{Schawinski_2009} identified 204 blue E/S0 galaxies using the Galaxy Zoo classification, which has blue $u-r$ colours that are significantly bluer than the red sequence and are well within the blue cloud in the optical colour-magnitude diagram occupied by star-forming galaxies. The redshifts of these galaxies are 0.02 $< z <$ 0.05, and the luminosities are greater than L$\star$. They are found to be in lower-density  environments than red sequence early-type galaxies and make up $\sim$ 6$\%$ of the low-redshift early-type galaxy population. Based on an analysis using emission line diagnostic diagrams, 25 $\%$ of these galaxies are actively star forming, 25 $\%$ host both star formation and an AGN, 12 $\%$ have an AGN, and 38 $\%$ show no strong emission lines that could be classified. With star formation rates ranging from 0.5 to 50 M$\odot$/yr, the star-forming blue E/S0 are found to be hosting intense, spatially extended star formation. We are interested in understanding the formation of the star-forming population of blue elliptical galaxies. Blue S0 galaxies can have different formation scenarios.\\

We used deeper optical imaging data to investigate whether the star-forming blue elliptical galaxies share a common origin with NGC 7252. We created a $g-r$ color map and measured the $g-r$ colours of selected merger features around the galaxy. We found  them to be of blue colours with a median $g-r$ = 0.52. The very central region of the galaxy has blue colors as well (g-r $\simeq$ 0.42) and is coincident with a star-forming disc revealed by HST \citep{Whitmore_1993} and UVIT far and near ultraviolet imaging \citep{George_2018b}. The main body of the galaxy has an elliptical morphology, which means that if it were detected without the tidal features, it would likely be classified as a blue elliptical galaxy.  We note that significant neutral hydrogen (4.5 $\times$ 10$^9$ M$\odot$) and molecular hydrogen (3.5 $\times$ 10$^9$ M$\odot$) gas is detected, which indicates that a gas-rich wet merger scenario is responsible for the formation of the merger remnant \citep{Hibbard_1994,Wang_1992}. 
The blue elliptical galaxies discussed in \citet{Schawinski_2009} have $>$ L$\star$ luminosity and therefore have a similar stellar mass as NGC 7252.\\

We explored the appearance of NGC 7252 at higher redshift with the likely detection of low surface brightness features from the merger. Figure~\ref{figure:fig3} shows that by redshift 0.7, the galaxy would not have detectable merger features in shallow surveys. This suggests that a merger remnant like NGC 7252 residing at z $\sim$ 0.7 would be morphologically classified as a blue elliptical galaxy. This is further demonstrated with the position of the galaxy main body on the grid of changing surface brightness for different redshifts between $0>z>1$ in Figure~\ref{figure:fig4}. The surface brightness of the main body of NGC 7252, shown with a black point, should be observed as an elliptical galaxy with a blue colour up to z$\sim$1. The merger features seen at the outskirts shown with colored points will disappear faster for shallow surveys such as the SDSS by z $\sim$ 0.1, but will appear for deeper surveys such as DECaLS till z $\sim$ 0.6. We note that this plot  only applies to sources with a flat spectral energy distribution, for which bandpass shifting plays a very negligible role.\\

We present here an idealised scenario for a merger remnant galaxy in the nearby Universe (z=0.0159) as it would appear at redshifts up to z $\sim$ 1. The galaxy can be more compact at high redshifts in wide-field imaging, which facilitates a location in the morphological classification as a blue elliptical. We did not consider the size reduction for elliptical galaxies that we observe at higher redshifts \citep{Trujillo_2007}. This effect is more prominent for massive elliptical galaxies, and minor mergers are likely responsible for galaxies that systematically increase their size at low redshifts \citep{Trujillo_2011}. We note that the  major merger rate decreases  towards lower redshifts, which can explain the low fraction of blue elliptical galaxies \citep{Lotz_2011}.\\

Features around merger remnant galaxies can disappear with time since the merger \citep{Ji_2014}. The NGC 7252 merger features are seen very clearly, implying a recent ($<$ 700 Myr) merger event. We note that almost all blue elliptical galaxies from \citet{Schawinski_2009}, although at varying levels, host features that are indicative of recent mergers revealed from a structural analysis and deep-imaging data \citep{George_2015,George_2017,George_2023}. The blue elliptical galaxies reported in the shallow SDSS imaging may be seen for longer time since the merger compared to the case of NGC 7252. Ongoing and future wide-field deep optical surveys (DES, DECaLS, Euclid, and LSST) will likely detect more blue elliptical galaxies at higher redshifts, and will need adequate surface
brightness sensitivity to reveal features that are indicative of recent merger activity.

\section{Summary}

Blue elliptical galaxies are interesting systems for understanding galaxy formation and evolution. We demonstrated based on deep-imaging data that the main body of the post-merger galaxy NGC 7252 can appear as a blue elliptical galaxy if it is observed at higher redshifts. With the evolution of the stellar population in the main body and the galaxy outskirts, the galaxy will most likely evolve into a normal elliptical galaxy with a red colour, hosting evolved stars. We argue that the blue elliptical galaxies 
 found from shallow imaging surveys may be post-merger systems, with the merger features going beyond the detection limit of the surveys.

\begin{acknowledgements}
The Legacy Surveys consist of three individual and complementary projects: the Dark Energy Camera Legacy Survey (DECaLS; Proposal ID 2014B-0404; PIs: David Schlegel and Arjun Dey), the Beijing-Arizona Sky Survey (BASS; NOAO Prop. ID 2015A-0801; PIs: Zhou Xu and Xiaohui Fan), and the Mayall z-band Legacy Survey (MzLS; Prop. ID 2016A-0453; PI: Arjun Dey). DECaLS, BASS and MzLS together include data obtained, respectively, at the Blanco telescope, Cerro Tololo Inter-American Observatory, NSF’s NOIRLab; the Bok telescope, Steward Observatory, University of Arizona; and the Mayall telescope, Kitt Peak National Observatory, NOIRLab. Pipeline processing and analyses of the data were supported by NOIRLab and the Lawrence Berkeley National Laboratory (LBNL). The Legacy Surveys project is honored to be permitted to conduct astronomical research on Iolkam Du’ag (Kitt Peak), a mountain with particular significance to the Tohono O’odham Nation.
NOIRLab is operated by the Association of Universities for Research in Astronomy (AURA) under a cooperative agreement with the National Science Foundation. LBNL is managed by the Regents of the University of California under contract to the U.S. Department of Energy.
This project used data obtained with the Dark Energy Camera (DECam), which was constructed by the Dark Energy Survey (DES) collaboration. Funding for the DES Projects has been provided by the U.S. Department of Energy, the U.S. National Science Foundation, the Ministry of Science and Education of Spain, the Science and Technology Facilities Council of the United Kingdom, the Higher Education Funding Council for England, the National Center for Supercomputing Applications at the University of Illinois at Urbana-Champaign, the Kavli Institute of Cosmological Physics at the University of Chicago, Center for Cosmology and Astro-Particle Physics at the Ohio State University, the Mitchell Institute for Fundamental Physics and Astronomy at Texas A\&M University, Financiadora de Estudos e Projetos, Fundacao Carlos Chagas Filho de Amparo, Financiadora de Estudos e Projetos, Fundacao Carlos Chagas Filho de Amparo a Pesquisa do Estado do Rio de Janeiro, Conselho Nacional de Desenvolvimento Cientifico e Tecnologico and the Ministerio da Ciencia, Tecnologia e Inovacao, the Deutsche Forschungsgemeinschaft and the Collaborating Institutions in the Dark Energy Survey. The Collaborating Institutions are Argonne National Laboratory, the University of California at Santa Cruz, the University of Cambridge, Centro de Investigaciones Energeticas, Medioambientales y Tecnologicas-Madrid, the University of Chicago, University College London, the DES-Brazil Consortium, the University of Edinburgh, the Eidgenossische Technische Hochschule (ETH) Zurich, Fermi National Accelerator Laboratory, the University of Illinois at Urbana-Champaign, the Institut de Ciencies de l’Espai (IEEC/CSIC), the Institut de Fisica d’Altes Energies, Lawrence Berkeley National Laboratory, the Ludwig Maximilians Universitat Munchen and the associated Excellence Cluster Universe, the University of Michigan, NSF’s NOIRLab, the University of Nottingham, the Ohio State University, the University of Pennsylvania, the University of Portsmouth, SLAC National Accelerator Laboratory, Stanford University, the University of Sussex, and Texas A\&M University.
BASS is a key project of the Telescope Access Program (TAP), which has been funded by the National Astronomical Observatories of China, the Chinese Academy of Sciences (the Strategic Priority Research Program “The Emergence of Cosmological Structures” Grant  XDB09000000), and the Special Fund for Astronomy from the Ministry of Finance. The BASS is also supported by the External Cooperation Program of Chinese Academy of Sciences (Grant  114A11KYSB20160057), and Chinese National Natural Science Foundation (Grant  12120101003,  11433005).
The Legacy Survey team makes use of data products from the Near-Earth Object Wide-field Infrared Survey Explorer (NEOWISE), which is a project of the Jet Propulsion Laboratory/California Institute of Technology. NEOWISE is funded by the National Aeronautics and Space Administration.
The Legacy Surveys imaging of the DESI footprint is supported by the Director, Office of Science, Office of High Energy Physics of the U.S. Department of Energy under Contract No. DE-AC02-05CH1123, by the National Energy Research Scientific Computing Center, a DOE Office of Science User Facility under the same contract; and by the U.S. National Science Foundation, Division of Astronomical Sciences under Contract No. AST-0950945 to NOAO.

\end{acknowledgements}


\end{document}